\documentclass[twoside,fleqn]{article}
\usepackage{latexsym,espcrc2,epsf}

\title{Use of even Grassmann variables to construct effective actions
for mesons} 
\author{
  Sergio Caracciolo \address{Scuola Normale Superiore and INFN,
      Sezione di Pisa,
      I-56100 Pisa, ITALIA},
 $\!$and
  Fabrizio Palumbo ~\thanks{This work has been partially 
  supported by EEC under TMR contract ERB FMRX-CT96-0045}
  \address{INFN -- Laboratori Nazionali di Frascati,
  P.~O.~Box 13, I-00044 Frascati, ITALIA}  
}

\begin{document}
\newcommand{\be}{\begin{equation}}
\newcommand{\ee}{\end{equation}}
\newcommand{\<}{\langle}
\renewcommand{\>}{\rangle}
\newcommand{\reff}[1]{(\ref{#1})}

\begin{abstract}
In the framework of an approach to bosonization based on the use of 
fermionic composites as fundamental variables, a quadratic
action in even Grassmann variables with the quantum numbers of the pions 
 has been constructed. It includes the $\sigma$-field in order to be invariant
under
$[SU(2)]_L\otimes[SU(2)]_R$ transformations over the quarks. This action
 exhibits the Goldstone phenomenon reducing its symmetry to
the O(3) isospin invariance.
The model has been investigated in the Stratonovitch--Hubbard
representation, in which form it is reminiscent of the
Gell-Mann-L\'{e}vy model. By the 
 saddle point method a renormalizable expansion in inverse powers of the
 index of nilpotency of the mesonic fields (which is 24), is generated.
The way it might be used in a new perturbative approach to QCD is outlined.
\end{abstract}

\maketitle

\section{Introduction}\label{introduction}

Effective Grassmann actions for even composites are in general polynomials of degree
equal to the index of nilpotency $\Omega$ of the composites~\cite{Palu}.  
QCD in strong coupling~\cite{Kawa,Barbour,Rossi}, the model defined 
in Ref.~\cite{Palu94} and the pairing model~\cite{Barb}  provide 
already examples where the action shows one and the same peculiar feature, 
namely the wave operator becomes static in the formal continuum limit, 
but, nonetheless it gives rise to a non trivial propagator. 
We have
considered a lattice action of this type for the pion, invariant 
under $[SU(2)]_L\otimes [SU(2)]_R$ transformations over the quarks. 
This model exhibits the  Goldstone phenomenon by reducing its symmetry to the 
$O(3)$ isospin invariance. In the presence of an explicit breaking  a squared mass for the pion is generated proportional to the 
breaking parameter.

\section{A quadratic action for the pions}\label{sec3}

The composites with the quantum numbers of the pions are
\begin{equation} 
  \vec{\pi} = i\,a^{2}\,\overline{\lambda} \gamma_5 \vec{\tau} \lambda \label{phidef}.
\end{equation}
In the above definition the $\tau_{k}$'s are the Pauli matrices and the sum over
the colour, isospin and spin  indices $a, f$ and $\beta$ of the quark field
$\lambda^{a}_{f,\beta}$  is understood. The power of the lattice 
spacing $a$ has been introduced in order to give the fields the 
canonical dimension of a boson.
To formulate a chiral invariant model we need also the field
\be
  \sigma = a^{2}\,\overline{\lambda} \lambda.
\ee 
All these composites 
have~\cite{Palu} index of nilpotency $\Omega =  24$. 

The chiral transformations over the quarks
\be
\delta \lambda = {i\over 2} \gamma_{5} \vec{\tau}\cdot \vec{\alpha} 
\lambda, \quad 
\delta \overline{\lambda} = {i\over 2} \overline{\lambda} \gamma_{5} 
\vec{\tau}\cdot \vec{\alpha} 
\ee
induce $O(4)$-transformations over the mesons
\be
\delta \sigma = \vec{\alpha} \cdot \vec{\pi},\quad
\delta \vec{\pi} = - \vec{\alpha} \sigma.
\ee
A quadratic action with the above symmetry must therefore
be of the form
\begin{equation}
S_{C} = -a^{-2}  \left[ {1 \over 2} (\vec{\pi},A\vec{\pi})  + {1 \over 2}
(\sigma,A\sigma)  +  (m, \sigma) \right],\label{actionC2} 
\end{equation}
where a breaking term has been included. The scalar product is defined as 
\be
(f,g) =  a^{4}\,\sum_{x} f(x) g(x).
\ee
The powers of the lattice spacing $a$ have been introduced assuming 
the wave-operator $A$ to be dimensionless.

The  definition of $A$ that we will use is
\be
 A = {\rho^4 \over - a^{2} \Box + \rho^{2}}\, . 
\ee
which becomes, for small lattice spacing,
\be
A \sim \rho^{2} (\rho^{2} + a^{2} \Box).
\ee
This, as previously announced, is static.
The first two terms  in the 
action \reff{actionC2} have  dimension 6 and we are therefore 
free to add them to the  QCD action.
The symmetry breaking  term  has instead dimension 4    (indeed it is of 
the same form of the quark mass term already
present in the  QCD action). We assume
\be
m = {\sqrt{\Omega}\over \rho} a m_{\pi}^{2}
\ee
so that it formally vanishes in the continuum 
limit and it can  be added to the QCD action.

Let us make clear that the partition function is
\begin{equation}
Z_{C} = \int [d\overline{\lambda} d\lambda]\, \exp(-S_{C}). 
\end{equation}

\section{Breaking of the chiral invariance}

In order to investigate our model we introduce the 
auxiliary fields $\vec{\chi}$ and $\chi_{0}$ by the Stratonovich--Hubbard 
transform.
The partition function can then be written 
\be 
   Z_{C} =  {1\over ( \det A)^{2}} \int \left[{d\vec{\chi}\over\sqrt{2 \pi}}\right]  
             \left[{d\chi_{0}\over\sqrt{2 \pi}}\right] 
 \exp (-S_{\chi})
\ee
where
\begin{eqnarray}
\lefteqn{S_{\chi} = {1\over 2 a^{2}} \left[ (\vec{\chi},A^{-1}\vec{\chi})  
+ (\chi_{0},A^{-1} \chi_{0}) \right] } \nonumber \\
& &-{\Omega\over 2 a^{4}} \left(1, \ln
D\left(\vec{\chi},\chi_{0}\right) \right)
\end{eqnarray}
and
\be
   D(\vec{\chi},\chi_{0}) = a^{2} \left[ \left( m+\chi_{0} 
   \right)^{2} + \vec{\chi}^{2} \right] .
\ee
This action is $O(4)$ invariant for vanishing quark mass and reminds us of the 
Gell-Mann--L\'{e}vy model~\cite{GML}.

 Since $\Omega$ is a rather large number we can  apply the saddle-point method 
and evaluate the partition function as a series in inverse powers of this
parameter. 
The minimum of $S_{\chi}$ is achieved for $\vec{\chi}=0$ and
\be
 \overline{\chi_{0}} = {-m a \pm \sqrt{m^{2}a^{2}+4 \Omega \rho^2} \over 2 a},
\ee
where the $+$ (respectively $-$) sign has to be chosen when $m>0$ 
(respectively $m<0$), which we shall assume to be the case. 

Now we assume 
 \begin{equation}
(am)^2<< 4\Omega \rho^2 \label{ine}
\end{equation}
so that 
\be
 \overline{\chi_{0}} \approx  a^{-1} [\sqrt{\Omega}  \rho - {1 \over 2} a m].  
\ee
The second derivatives of $S_{\chi}$ at the minimum are
\begin{eqnarray}
\lefteqn{\left.{\partial^2 S_\chi \over \partial \chi_0(x) \partial \chi_0(y)}
\right|_{\chi=\bar{\chi}} =
   a^2 \left[ A^{-1}_{xy}+ \delta_{xy}{\Omega \over { D}}  \right] }
   \nonumber \\ & & \hspace{1.8cm} =
  {a^4 \over {\rho^4}} \left( -\Box +  { 2 \rho^{2}\over{ a^2}} -
 m {\rho  \over a \sqrt{\Omega}}\right)  
\nonumber\\
\lefteqn{\left.{\partial^2 S_\chi \over \partial \chi_h(x) \partial \chi_k(y)}
\right|_{\chi=\bar{\chi}} =
 \delta_{hk}    a^2 \left[ A^{-1}_{xy}- \delta_{xy}{\Omega \over {
D}}  \right] }
   \nonumber \\  & & \hspace{1.8cm} =
   \delta_{hk} {a^4 \over {\rho^4}} \left( -\Box  + m {\rho \over a \sqrt{\Omega}} 
\right).  
\end{eqnarray}
The propagator of the pion field to leading order 
\be
<\pi_h(x) \pi_k(y)> =   {1\over a^{4}} \,\left({ 1 \over { -\Box+
m_{\pi}^2}} \right)_{x,y}
\ee
turns out to have the canonical form. 

The $\sigma$-field  acquires the nonvanishing expectation value  
\begin{equation}
<\sigma>= {1\over V} {\partial\over \partial (a^2m)} \log Z_{C}
 \approx a^{-1} \rho^{-1}
\sqrt{\Omega} \label{vms} 
\end{equation}
and its mass is divergent according to
\be
m_{\sigma}^2= 2 a^{-2} \rho^2 -m_{\pi}^2 = 2 {M_{3}\over a}-m_{\pi}^2 . \label{ms}
\ee
It is a consequence of the inequality~\reff{ine} that $m_\sigma >>
m_\pi$.

\section{The $\Omega$-expansion}

In order to formulate the $\Omega$-expansion it is convenient  to introduce the fields
which correspond to rescaled fluctuations
\be
\theta_0  =   \rho^{-2 }( \chi_0 - \bar{\chi}_0 ),
\quad
\theta_k  =   \rho^{-2} \chi_k.
\ee
In terms of these fields the function $D$ takes the form
\be
D=  C^2 F
\ee
where
\begin{eqnarray}
C &=&\sqrt{\Omega \rho^2 + { 1\over 4} a^2m^2 } + { 1\over 2} am \approx \sqrt{\Omega}
\rho+ { 1 \over 2} am
\nonumber\\
F &=&  1+ {\rho^{2} \over C} 2a \theta_0  + {\rho^{4}\over C^2} a^2 
( \theta_0^2 + {\vec{\theta}}^2 ).
\end{eqnarray}
 By expanding the $\ln D $ we  rewrite the action $S_{\chi}$ as a series
\begin{equation}
S_{\chi}= \sum_{n=2}^{\infty}  S^{(n)}.
\end{equation}
The term $S^{(n)}$, for $n>2$, is a homogeneous polynomial of degree $n$ in the
$\theta$-fields   proportional to $ \rho^{2n} a^n / {C^n}$. The first three terms are
\begin{eqnarray}
S^{(2)} &=& a^4 \sum_x {1 \over2} \vec{\theta} \left(-\Box+m_{\pi}^2\right)
\vec{\theta} \nonumber\\
& & +{1\over2} \theta_0 \left(-\Box+ m_{\sigma}^2\right) \theta_0 \nonumber\\
S^{(3)} &=&  {\rho^6 \over C^3} a^4\sum_x { 1\over a}\left(  { 2\over 3} \theta_0^3
 -2 \theta_0 \vec{\theta}^2  \right) \nonumber\\
S^{(4)} &=& { \rho^8 \over C^4} a^4\sum_x  -{1\over 2} \theta_0^4 
-{1\over 2} (\vec{\theta}^2)^2 + 3 \theta_0^2  \vec{\theta}^2 .
\end{eqnarray}
$S^{(n)}$ turns out to be of the order $\Omega^{-{n\over2}}$, for $n>2$.
We actually have then an expansion in inverse powers of $\sqrt{\Omega}$, but the first
correction is of order $\Omega^{-{3\over 2}}$. 

We can now investigate  the renormalizability of this expansion.
This requires $\rho^2 / C$ not to diverge in the
continuum limit,
which is, indeed, the case.

We should mention that by a different dependence of $\rho$ and $m$ on 
the lattice spacing we can get a {\em truly 
free action} for the pions, all the interaction terms giving vanishing 
contribution to the $n$-point functions. The resulting action, 
however, cannot be added to the QCD action as an irrelevant operator, 
because it is not accompanied by the necessary powers of the lattice 
spacing. Therefore it cannot be used to set up
a new perturbative approach to QCD, which is our main motivation.
In such a case one can regard such a result
 as the construction of a simple model, whose main
ingredient is the compositness of the mesonic fields.

\end{document}